\documentclass[12pt,letterpaper]{article}
\usepackage{amsmath,amssymb,pgf,pgfarrows,pgfnodes,float,appendix, hyperref}
\usepackage{graphicx}
\usepackage{subfigure}
\usepackage[margin=0.9in]{geometry}

\newcommand{\be}{\begin{equation}}
\newcommand{\ee}{\end{equation}}
\newcommand{\bea}{\begin{eqnarray}}
\newcommand{\eea}{\end{eqnarray}}

\def\S{{\cal S}}

\title{{\rm\footnotesize \qquad \qquad \qquad \qquad \qquad \ \qquad \qquad \qquad \ \ \ \ \ \                  RUNHETC-2019-33, UTTG-27-19 }\vskip.5in   Systematic Resummation of the Large N expansion of Vector Models: Application to the Hubbard model and $2 + 1 $ dimensional QED}
\author{Tom Banks\\
Department of Physics and NHETC\\
Rutgers University, Piscataway, NJ 08854\\
E-mail: \href{mailto:banks@physics.rutgers.edu}{banks@physics.rutgers.edu}
\\
\\
}
\date{}
\begin{document}
\maketitle

\begin{abstract}
We introduce a hierarchy of closed equations for charge density correlation functions in the Hubbard model and $2 + 1$ dimensional QED.  Each step in the hierarchy can be considered a large $N$  truncation of an exact, but infinite set of equations relating all $k-$point charge correlators.  $N$ is the number of fermion spin components.  Each step in the hierarchy sums up an infinite number of large $N$ diagrams, including all diagrams up to some fixed order, for $k$ point functions with $k \leq K$.  Higher point functions are replaced with their leading large $N$ behavior.  The simplest truncation gives a closed nonlinear equation for the $2$ point function. 
\end{abstract}

\section{Introduction}

The Hubbard model, with Hamiltonian \begin{equation} H = \sum_{ij} t_{ij} \psi_i^{\dagger\ a} \psi_j^a + g^2 \sum_i (\psi_i^{\dagger\ a} \psi_i^a )^2 , \end{equation} is the universal workhorse of modern condensed matter theory.  One incorporates a lot of the physics of a given material into the tight binding coefficients $t_{ij}$ which determine the crystal structure of the ground state, and the rest is supposed to be well approximated by a choice of the on site Coulomb repulsion $g^2$.   In the simplest Hubbard models the spin index $a$ takes on two values, representing the spin of electrons.  Some materials require more complicated multi-band Hubbard models.  The basic idea behind the Hubbard model is that Coulomb forces are screened, with a screening length shorter than the lattice spacing.  With the exception of phonons, low energy excitations are assumed to be excitations of this low energy electron gas.

The imaginary time Lagrangian of the Hubbard model may we rewritten as 
\begin{equation} {\cal L} =\psi_i^{\dagger\ a}  \sum_{ij} [ (\partial_t + i\sigma) \delta_{ij} + t_{ij} ]\psi_j^a + \frac{\sigma^2}{2g^2} . \end{equation}  Integrating over the Hubbard-Stratonovich field $\sigma$ in the Feynman path integral, we recover the original Hamiltonian.  The large $N$ expansion devolves from the observation that in the limit of a large number of field components, if $g^2 = \frac{\lambda}{N}$, then the effective action for the HS field is proportional to $N$.  The stationary point of this action, constant in space and time is given by the equation
\begin{equation} \sigma_0  = \lambda G(t,x; t,x) ,\end{equation} where the Green's function $G$ is the inverse of the operator $(\partial_t + i \sigma_0) \delta_{ij} + t_{ij}$, with thermal boundary conditions. 

\begin{figure}[h!]
\begin{center}
  \includegraphics[width=12cm]{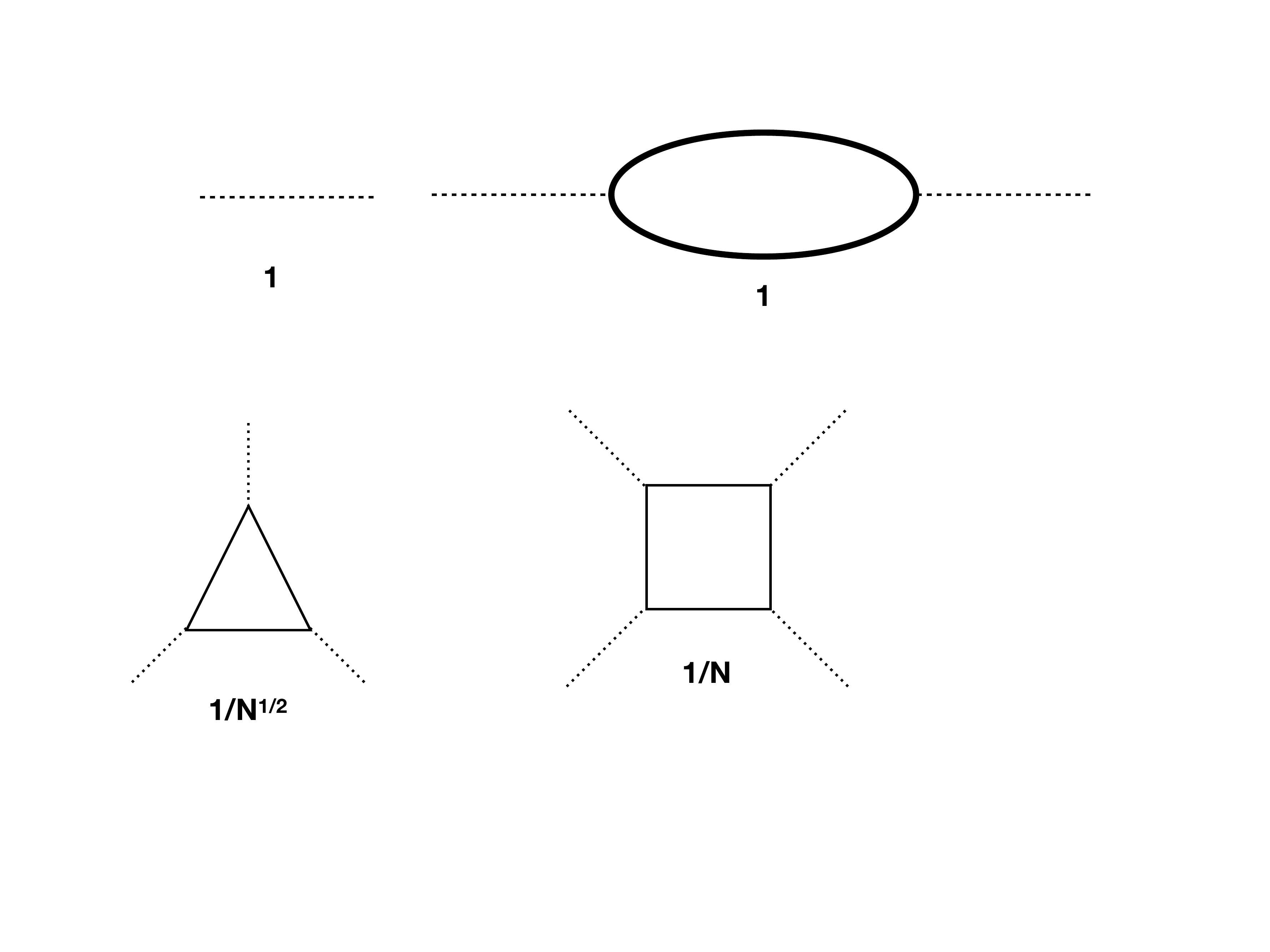}
\end{center}
\vspace{-1.4cm}
\caption{ Large N Vertices.\label{LargeNvert}}
\end{figure}

If we define $\sigma = \sigma_0 + N^{-1/2} s$, then the large $N$ expansion is a perturbative expansion for correlation functions of $s$.  The inverse propagator is $\lambda^{-1} \delta_{ij} + \Pi_{ij}$, where $\Pi$ is the one fermion loop vacuum polarization diagram. The vertices $S_k$ for $k \geq 3$ are similar single fermion loop diagrams, and proportional to $N^{(2 - k)/2} $.  See Figure \ref{LargeNvert}.  

The functional integral formula for the generating functional of connected correlators of $s$ leads to the exact equation
\begin{equation} \partial_z W = \int dt\ \sum_x [\frac{\delta^2 W}{\delta J(t,x) \delta J(t,x)} + \frac{\delta W}{\delta J(t,x)}\frac{\delta W}{\delta J(t,x)}] , \end{equation} where $z \equiv \lambda^{-1}$ .   One particle irreducible (1PI) vertices $\Gamma_k$ for the $s$ field are defined perturbatively by the sum of all Feynman diagrams in the $1/N$ expansion, which cannot be cut by cutting a single propagator line.  It's clear diagramatically that general connected diagrams are sums over "trees" whose branches are the full propagator and whose "k-crotches" are the $\Gamma_k$.  It's well known that the generating functional for the $\Gamma_k$ is the functional Legendre transform of that for the connected correlators $W_k$ .  
\begin{equation} \Gamma[s_c] + \int\ \sum s_c (t,i) J(t,i) = W[J] , \end{equation} where 
\begin{equation} \frac{\delta \Gamma}{\delta s_c (t,i)} = - J(t,i) . \end{equation}  The second equation implies that
\begin{equation} \int dt^{\prime} \sum_j \frac{\delta^2 \Gamma}{\delta s_c (t,i) \delta s_c (t^{\prime},j)}  \frac{\delta^2 W}{\delta J (u,k) \delta J (t^{\prime},j)} = - \delta_{ik} \delta (t - u) . \end{equation}
\begin{equation} \frac{\delta^2 \Gamma}{\delta s_c (t,i) \delta s_c (t^{\prime},j)} = - \frac{\delta J (t,i)}{ \delta s_c (t^{\prime},j)}  . \end{equation}

We can use these relations to get an exact hierarchy of equations for the $\Gamma_k$
\begin{equation} (1 - k) \partial_z \Gamma_k (\ldots ) = \int dt\ \sum_x [W_{k + 2}^T (\ldots , t, x ; t,x)]+ \delta_{k2} \delta_{ij} \delta (u - v)  . \end{equation} In this equations $\ldots$ represents all the arguments on which $\Gamma_k$ depends.  In the case $k = 2$, these arguments are 
the times $u,v$ and the lattice positions $i,j$.  The superscript $T$ on $W_{k + 2}$ implies that all of the external propagators in its tree expansion in terms of $\Gamma_{k + 2}$ and lower point vertices are "amputated".  The propagators on the external legs in the tree expansion are dropped, while those inside the integral above are retained. Using the tree expansion, these equations 
form an infinite hierarchy of nonlinear relations between the $\Gamma_k$.  

As a differential equation in $z$, these relations need to be supplemented by a boundary condition.  Since the large $z$ limit is perturbative we take the large $z$ boundary condition to be that each $\Gamma_k$ is given by its large $N$ limit. For $\Gamma_2$ this is given by two diagrams in large $z$  perturbation theory, while for higher point functions it just gives the leading perturbative diagram.  In this prescription, we keep the first subleading term in powers of $z^{-1}$ in the two point function.   In the Homogeneous Electron Fluid, where a similar hierarchy was first introduced\cite{tbheg}, one {\it must} do that in order to avoid infrared divergences that give rise to Debye screening.  For the Hubbard model, keeping the second term is optional.  We hope that doing so will improve the accuracy of the approximation.  
\section{Large $N$ Truncation of the Hierarchy}

The equation for $\partial_z \Gamma_k$ involves $\Gamma_{l \leq k}$, $\Gamma_{k + 1}$ and $\Gamma_{k+2}$.  The $k-$th truncation step consists of approximating $\Gamma_{k + 1,2}$ by their leading order, one fermion loop, approximations.  The resulting set of equations captures the large $N$ expansion exactly up to order $N^{(1 - k)/2}$.  At each step it is a closed set of equations for the 1PI correlators, $\Gamma_2 \ldots \Gamma_k$, which resums an infinite number of additional large $N$ terms.  For $k = 2$ we get a closed equation for the two point correlation function.  It sums up all Feynman graphs in the large $N$ expansion containing only $\Gamma_{3,4}$ vertices and approximates those vertices by $S_{3,4}$.  

The strategy for solving these equations numerically utilizes the standard iterative solution for systems of first order differential equations.  Consider a system
\begin{equation} \partial_z q^i = F^i (q) , \end{equation} with a boundary condition
\begin{equation} q^i (z_0) = q^i_0 . \end{equation} Then
\begin{equation} q^i (z) = q^i_0 - \int_z^{z_0}\ F^i (q (z)) . \end{equation}  Given an initial guess, $q^i_1 (z)$, for  the functions, this formula defines a mapping 
\begin{equation} q^i_{n + 1} (z) = q^i_0 - \int_z^{z_0}\ F^i (q_n (z)) . \end{equation} For rather general choices of $F^i$ , this defines {\it a contraction mapping on a complete metric space} of functions $q^i (z)$.  That is, one can equip the function space with a notion of distance between functions, and show that the mapping always decreases the distance.  This proves existence and uniqueness of the solution, and provides a constructive method for finding it.

The abstract indices $i$ in the above paragraph represent the values of the correlations $\Gamma_1 \ldots \Gamma_k$ at all points in time and all lattice points.  If we work in finite volume and restrict time to a large finite interval $[-T,T]$ then the functions $F^i$ in our equations are finite sums of rational functions of the variables, and the conditions for a contraction mapping are preserved.  The possible failure of these conditions in the limit of large space-time volume is related to phase transitions and presumably occurs only at discrete values of $z$.

The numerical strategy indicated by these theorems is to turn the differential equation into a difference equation, by discretizing the $z$ axis, and choose a value of $z_0$ large enough to trust the leading large $N$ approximation to the $\Gamma_l$, for $l \leq k$, which appear as unknowns in the $k-$th truncation of the hierarchy.  We then use the difference equation to obtain approximate solutions for these $\Gamma_l$ at smaller values of $z$.  Continuous phase transitions will be located by finding values of $z$ where the numerical integrations begin to diverge.  

\section{The Two Point Truncation at Large $N$}
The explicit form of the leading truncation is,

\begin{equation} 
\begin{aligned}
(-1) \partial_z \Gamma_2 (\omega_q, q; - \omega_q, - q ) =  \int d\omega dp\  \Gamma_2^{-1} (\omega, p) \Gamma_2^{-1} (- \omega, - p) [ \S_4 (\omega, p; - \omega, - p; \omega_q, q; -\omega_q, - q)\\    - S_3 (\omega, p; - \omega, - p ; 0,0; \omega_q , q; - \omega_q , - q)\Gamma_2^{-1} (0,0) S_3 (0,0; \omega_q, q; -\omega_q, - q) \\        - 2 S_3 (\omega, p; - (\omega + \omega_q), - p - q ; \omega_q , q) \Gamma_2^{-1} (\omega + \omega_q , p + q) S_3 (\omega, p;  (\omega + \omega_q),  p - q ; -\omega_q , -q) ] + 1   . 
\end{aligned}
\end{equation}
  This should be supplemented with the large $z$ boundary condition
\begin{equation} \Gamma_2 \rightarrow z \delta_{ij} \delta (u - v) + \Pi (i - j, u - v) , \end{equation} where $\Pi$ is the one fermion loop two point function.  

The utility of this equation depends on finding a good numerical approximation to $S_3$ and $S_4$.  The one loop frequency integrals can be done by contour integration, but one must still do complicated lattice sums  to extract these functions.

\section{More General Models}

The previous analysis can be generalized to a much more general class of models involving $N$ component bosons or fermions coupled to a singlet boson with a quadratic action.  Starting with a Euclidean lattice model we write
\begin{equation} S = N g \sum_{ij}  \chi_i \chi_j  K_{ij} + \sum_{ij} \bar{\psi}_i^a \psi_j^a t_{ij}+ \sum_i \chi_i \bar{\psi}_i^a \psi_i^a . \end{equation}  Integrating over the non-singlet fields we get
\begin{equation} S_{eff} = N [ g \sum_{ij}  \chi_i \chi_j  K_{ij}  \pm {\rm tr\ ln\ } (t_{ij} + \delta_{ij} \chi_i ) ] \end{equation}
\begin{equation} = [ g \sum_{ij}  \chi_i \chi_j  K_{ij}  \pm N {\rm tr\ ln\ } (t_{ij} + N^{-1/2} \delta_{ij} \chi_i) ] .\end{equation}  The $\pm$ sign is the Fermi/Bose alternative for the non-singlet fields.

As above, this leads to the exact functional equation
\begin{equation} \partial_g W[J] = \sum_{ij} K_{ij} [W_2 (i,j) + W_1 (i) W_1 (j)] , \end{equation} which translates into 
\begin{equation} (1 - n) \partial_g \Gamma (i_1 \ldots i_n) = \sum_{ij} K_{ij} [W_{n + 2}^T (i,j, i_1 \ldots i_n ) + \delta_{n2} \delta_{ij} ] , \end{equation} for $n$ point 1PI correlators of $\chi$.
We also have the large $N$ scaling $\Gamma_n \sim N^{1 - n/2} $.  
Writing $W_{n + 2}$ as a tree expansion in 1PI correlators, we get a hierarchy of non-linear equations for the $\Gamma_n$.  We supplement these with the large $g$ boundary condition that each $\Gamma_n$ with $n > 2$ approaches its large $N$ limit, while $\Gamma_2$ retains its single fermion loop contribution, and contains terms of both order $1$ and order $1/N$.  This can be important when the single loop dominates the tree level term at low momentum, as often happens in models with infrared issues.
Note by the way that the difference between bosonic and fermion models in these equations is all contained in the {\it signs} of the vertices $S_k$ with $k > K$.

The systematic resummation scheme consists of truncating this hierarchy by approximating $\Gamma_n $ with $n > K \geq 3$ by their large $N$ limits, which are explicitly calculable.  For each choice of $K$, the resulting equations sum up an infinite number of large $N$ diagrams for $\Gamma_n$ with $n \leq K$ including all diagrams up to order $N^{- K}$.   The equation for the two point function, when $K = 3$, has the form
\begin{equation}
\begin{aligned}  
(-) \partial_g \Gamma_2 (i_1, i_2) = N^{-1} \sum_{ijkl} K_{ij} \bigl{(} \Gamma_2^{-1} (i, k) \Gamma_2^{-1} (j, l)\bigl{[}S_{4} (k,l, i_1, i_2 ) - S_3 (k,l,r) \Gamma_2^{-1} (r,s) S_3 (s,i_1,i_2 ) - \\ S_3 (k,i_1,r ) S_3 (l,i_2,s ) \Gamma_2^{-1} (r,s)  - S_3 (l,i_1,r ) S_3 (k,i_2,s ) \Gamma_2^{-1} (r,s)\bigr{]}+  \delta_{i_1i}\delta_{i_2 j} \bigr{)} . 
\end{aligned}
\end{equation} 
In this equation we've factored out the explicit power of $N^{-1}$ on the right hand side, so $S_{3,4}$ should be evaluated as quantities of order $1$ in the large $N$ expansion.  The solution of this equation sums up all diagrams in the large $N$ expansion for the two point function of $\chi$, containing only three and four point one loop vertices, with no internal decoration.  
\section{Three dimensional QED}

The Lagrangian of this model is 
\begin{equation} {\cal L} = - \frac{N}{4 g^2} B_{\mu\nu}^2 + \bar{\Psi}_i (i\gamma^{\mu} \partial_{\mu}  - \gamma^{\mu}B_{\mu} - m ) \Psi_i .  \end{equation}  We work in Euclidean space with $\Psi_i$ transforming as $N$ copies of the $[2]$ of $SU(2)$.  Write $B_{\mu} = m A_{\mu} $, $\Psi_i = m \psi_i$ so that the action becomes
\begin{equation} S = \int d^3 x\ [- \frac{Nm}{4 g^2} A_{\mu\nu}^2 + \bar{\psi}_i (i\gamma^{\mu} \partial_{\mu}  - \gamma^{\mu}A_{\mu} - 1 ) \psi_i ].  \end{equation}

Rescaling $A_{\mu} = 1/\sqrt{N} a_{\mu}$ we get the effective action 
\begin{equation} S_{eff} [a] =  - \frac{m}{4 g^2} a_{\mu\nu}^2 + N {\rm Tr\ ln\ } (i\gamma^{\mu} (\partial_{\mu} - \frac{a_{\mu}}{\sqrt{N}}) - 1) . \end{equation}  
Define $z = m/g^2$.  Couple a source to the vector potential $\int j_{\mu} a^{\mu}$ and then derive an equation for the derivative of the functional integral with this effective action (plus a gauge fixing term) with respect to $x$.  The equation reads
\begin{equation} \partial_z Z[j] = \int \frac{d^3 p}{(2\pi)^3}\ D_{\mu\nu} (p) \frac{\delta^2 Z}{\delta j_{\mu} (p)\delta j^{\nu} (- p) } . \end{equation} Written in terms of 1PI correlation functions this equation is
\begin{equation} (n - 1) \partial_z \Gamma_{n\ A} (p_i) = \int \frac{d^3 p}{(2\pi)^3}\ [- W^T_{(n + 2)\ A} (p, - p , p_i)]  + 3\delta_{n2} \delta^3 (p_1 - p_2) D_{A_1\ A_2} (p_1). \end{equation}  For brevity of notation we have used the multi-index $A = (\mu_1 \ldots \mu_n ) $ which is symmetric under interchange of indices.  $D_{\mu\nu} (p)$ is the bare inverse photon propagator in the chosen gauge.

If we use the tree expansion of connected correlators in terms of 1PI correlators, this becomes a hierarchy of coupled non-linear integro differential equations for the $\Gamma_{n\ A}$.  To solve them, we use the large $z$ boundary condition that each correlator approaches its large $N$ limiting value.  For $n \geq 4$ this is just the single fermion loop diagram, which also dominates the large $x$ limit,  while for the two point 
function it contains both tree and one loop contributions to the large $x$ expansion.  

\begin{figure}[h!]
\begin{center}
  \includegraphics[width=12cm]{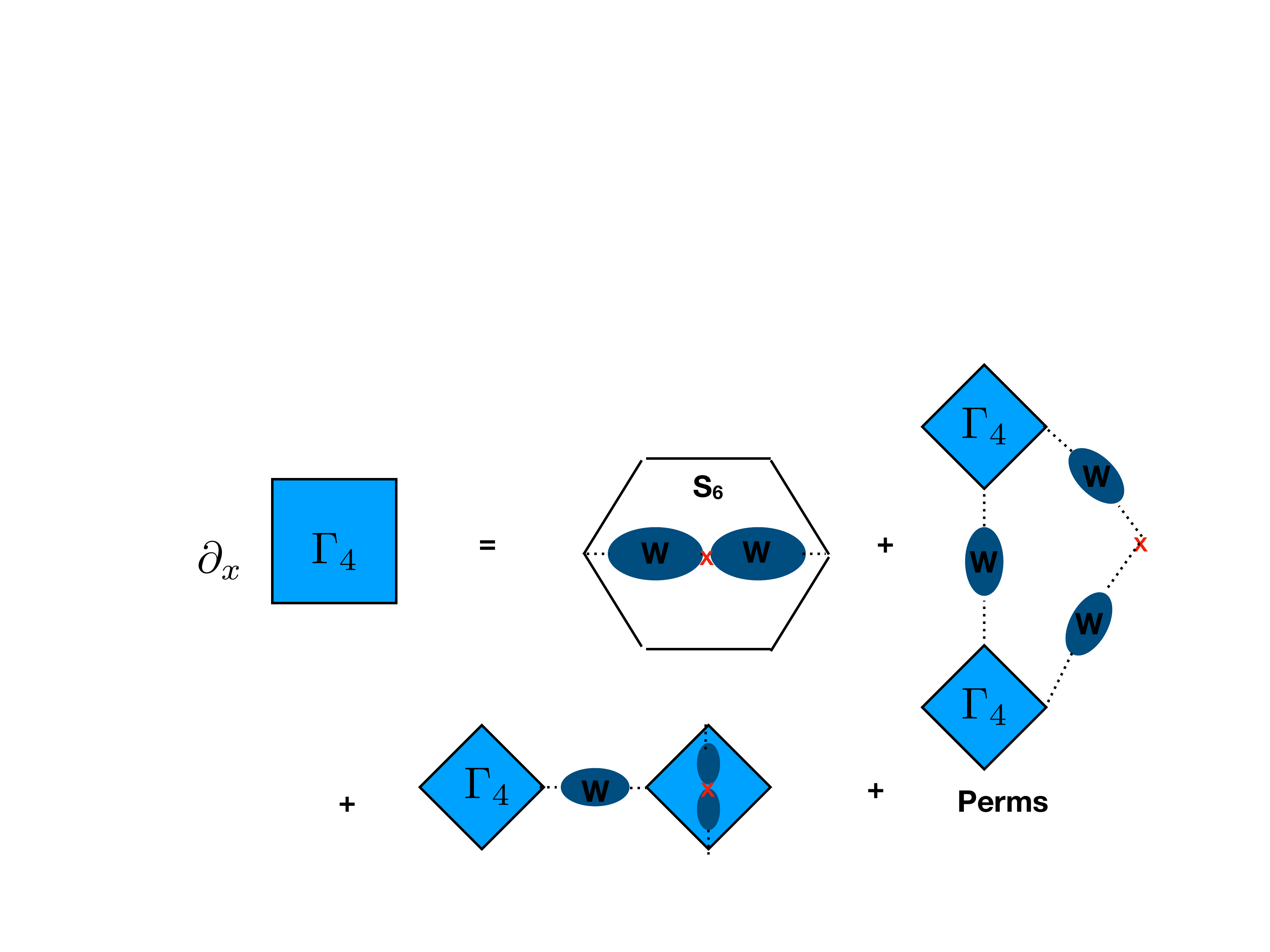}
\end{center}
\vspace{-1.4cm}
\caption{ Equation for 1PI Four Point Function in QED.\label{fourpoint}}
\end{figure}

We can now contemplate truncations of this hierarchy of equations at large $N$.  The 1PI $k$ point functions are all dominated by graphs with a single fermion loop and scale like $N^{1 - k/2}$.  This model has a charge conjugation symmetry, which restricts $k$ to be even.   As a consequence, the connected $4$ point function and connected $6$ point functions are given in terms of a single tree graph of 1PI functions.  The leading order truncation of the hierarchy is a closed equation for the 1PI $2$ point function
\begin{equation} \partial_z \Gamma_{\mu\nu} (q) = N^{-1} \int \frac{d^3 p}{(2\pi)^3} S^{(4)}_{\alpha\beta;\mu\nu} (p, - p, q, -q) G_{\alpha\rho} (p) G_{\beta\sigma} ( - p) D_{\rho\sigma} (p) + D_{\mu\nu} (q) . \end{equation}  Here $ - G$ is the matrix inverse of $\Gamma$.  $S^{(4)}$ is given by the usual (off shell) light by light scattering diagram.  At the next level of truncation we make the replacement $N^{-1}S^{(4)}_{\alpha\beta;\mu\nu} (p, - p, q, -q) \rightarrow \Gamma^{(4)}_{\alpha\beta;\mu\nu} (p, - p, q, -q)$ in this equation and add an additional equation for $\Gamma^{(4)}$, which is shown graphically in Figure \ref{fourpoint}.   

The integral equations of the $p-th$ truncation resum an infinite number of terms in the large $N$ expansion of the unknown functions including all terms of order less than or equal to $N^{-p}$.  Numerically, the strategy for solving them seems straightforward.  Write the $z$ derivative as a finite difference and start at a value of $x$ large enough that one trusts that the leading order large $N$ approximation is accurate.  This allows one to compute the integrals that give the value of the correlators at a slightly smaller value of $z$.  Rinse and repeat.  
It might also be useful to exploit the super-renormalizability of the interaction to do the high momentum part of the integral analytically.  In that case any slowness of the convergence of the numerical integration is likely to come from interesting IR physics.

The correlation functions we have defined are of course gauge dependent, but the standard Ward identities of abelian theories tell us how to extract gauge invariant information from them.

The massless limit of this theory $z \rightarrow 0$ is believed to have an interesting phase diagram\cite{appel} as a function of $N$.  Below a value $N_C \sim \frac{128}{3\pi^2} $ it has a spontaneous symmetry breaking phase with a massive fermion, while above that value it flows to a non-trivial conformal field theory.  Since our basic equations involve derivatives with respect to $z$, they have no meaning in the conformal field theory itself.  Rather, viewing non-zero $z$ as a relevant perturbation of the fixed point, an equation of the form
\begin{equation} \partial_z \Gamma_k (q)  \sim \int d^3 p\ D(p) W^T_{k+2} (p, - p; q) , \end{equation} determines the scaling dimension of the relevant operator.  If the dimension of the electromagnetic potential at $z = 0$ is $\Delta$ this equation fixes the dimension of the parameter $z$ to be $ 2 \Delta - 1$.  So we must have $2\Delta > 1$ in order for $z$ to be relevant.  It seems plausible that a non-perturbative analysis of the equations in this paper will lead to more detailed information about the conformal theory that exists for large $N$.

 \vskip.3in
\begin{center}

{\bf Acknowledgments } The work in this paper was partly supported by the U.S. Department of Energy under Grant DE-SC0010008.\\

\end{center}

\end{document}